\documentclass[aps,prl,preprintnumbers,twocolumn,groupedaddress,showpacs]{revtex4}

\usepackage{latexsym}
\usepackage{amsmath}
\usepackage{amsfonts}
\usepackage{graphicx}
\usepackage{axodraw}

\newcommand{\bq}{\begin{eqnarray}}
\newcommand{\eq}{\end{eqnarray}}

\newcommand{\eps}{\varepsilon}

\begin{document}

\preprint{MZ-TH/09-36}
\title{\boldmath{A simple formula for the infrared singular part of the integrand of one-loop QCD amplitudes}}

\author{Mohammad Assadsolimani, Sebastian Becker and Stefan Weinzierl}
\affiliation{Institut f{\"u}r Physik, Universit\"at Mainz, D-55099 Mainz, Germany}

\date{\today}

\begin{abstract}
We show that a well-known simple formula for the explicit infrared poles 
of one-loop QCD amplitudes has a corresponding simple counterpart in unintegrated form.
The unintegrated formula approximates the integrand of one-loop QCD amplitudes in all soft and collinear singular regions.
It thus defines a local counter-term for the infrared singularities and can be used as an ingredient for the numerical
calculation of one-loop amplitudes.
\end{abstract}

\pacs{11.15.Bt, 12.38.Bx, 13.87.-a}

\maketitle

%-----------------------------------------------------------
\section{Introduction}

The calculation of next-to-leading order (NLO) QCD corrections for multi-jet final states plays an important 
role for the experiments at the LHC.
For any NLO calculation there are two parts to be calculated: the real and the virtual corrections.
As far as the real corrections are concerned it has become a standard approach to first 
subtract out the infrared divergences \cite{Catani:1997vz,Phaf:2001gc,Catani:2002hc}
and to perform the phase space integration of the subtracted real correction term
numerically by Monte Carlo techniques.
The subtraction terms can be integrated and added back. The integrated form has a remarkable simple structure
and predicts the infrared poles of the renormalised one-loop amplitude.
The one-loop amplitude contributes to the virtual part and has in general ultraviolet, soft and collinear divergences.
It involves the integration over one unconstrained momentum (the loop momentum) in $D$ dimensions.
It is tempting to follow a similar path also for the virtual part: First subtract out all singular terms of ultraviolet and infrared 
origin from the one-loop amplitude and then perform the integration over the loop momentum together with the
phase space integration in four dimensions by deforming the loop integration contour into the complex plane.
The subtraction terms are then integrated analytically in $D$ dimensions and 
added back.
Such subtraction terms have been considered in \cite{Nagy:2003qn} and issues related to the required contour deformation have
been studied in \cite{Nagy:2006xy,Gong:2008ww}.
The unintegrated infrared subtraction terms in \cite{Nagy:2003qn} are determined graph by graph.
This limits the efficiency of the method. 
Inspired by recent work on the structure of infrared singularities of multi-loop amplitudes \cite{Becher:2009cu,Gardi:2009qi,Ferroglia:2009ep}
we show in this letter that the soft and collinear subtraction terms can be formulated at the level of amplitudes, 
without referring to individual
Feynman graphs. 
This is a significant simplification and opens the door to an efficient implementation based on recurrence relations \cite{Berends:1987me}.
The required ultraviolet subtraction terms have the form of propagator and vertex counter-terms and can be incorporated into
the recurrence relations. 
A possible choice for the ultraviolet subtraction terms is given in ref.~\cite{Nagy:2003qn}.

%-----------------------------------------------------------
\section{Notation}
\label{sect:notation}

Let us consider a one-loop QCD amplitude with $n$ external partons.
The full one-loop amplitude can be colour-decomposed into
primitive amplitudes:
\bq
 {\cal A}^{(1)} & = & \sum\limits_{j} C_j A^{(1)}_j.
\eq
The colour structures are denoted by $C_j$ while
the primitive amplitudes are denoted by $A^{(1)}_j$.
In the colour-flow basis \cite{'tHooft:1973jz,Maltoni:2002mq,Weinzierl:2005dd} the colour structures are linear combinations of monomials in Kronecker $\delta_{ij}$'s.
Primitive amplitudes are defined as a colour-stripped 
gauge-invariant set of Feynman diagrams with a fixed cyclic ordering of
the external partons and a definite routing 
of the external fermions lines through the diagram \cite{Bern:1994fz}.
We have collected some information on primitive amplitudes in an appendix.
It is convenient and sufficient to focus on primitive amplitudes.
In the following we drop the subscript $j$ and focus on a 
single primitive one-loop amplitude which we denote by $A^{(1)}$.
Since the cyclic ordering of the external partons is fixed, there are only $n$ different propagators occurring
in the loop integral.
We label the external momenta clockwise by $p_1$, $p_2$, ..., $p_n$ 
and define $q_i=p_1+p_2+...+p_i$, $k_i=k-q_i$.
We can write the bare primitive one-loop amplitude in Feynman gauge as
\bq
\label{starting_point}
 A^{(1)}_{bare} & = & \int \frac{d^Dk}{(2\pi)^D} 
 G^{(1)}_{bare},
\nonumber \\
 G^{(1)}_{bare} & = &
 P(k) \prod\limits_{i=1}^n \frac{1}{k_i^2 - m_i^2 + i \delta}.
\eq
$G^{(1)}_{bare}$ is the integrand of the bare one-loop amplitude.
$P(k)$ is a polynomial in the loop momentum $k$.
The $+i\delta$-prescription instructs us to deform -- if possible -- the integration contour into the complex plane
to avoid the poles at $k_i^2=m_i^2$.
If a deformation close to a pole is not possible, we say that the contour is pinched.
In this letter we restrict ourselves to non-exceptional external momenta.
Then the divergences of the one-loop amplitude related to a pinched contour are either due to soft
or collinear partons in the loop.
These divergences are regulated within dimensional regularisation by setting the number of space-time dimensions
equal to $D=4-2\eps$.
A primitive amplitude which has soft or collinear divergences must have at least one loop propagator which corresponds
to a gluon. An amplitude which just consists of a closed fermion loop does not have any infrared divergences.
We denote by $I_g$ the set of indices $i$, for which the 
propagator $i$ in the loop corresponds to a gluon.
If we take the subset of diagrams which have the gluon loop propagator $i$ and if we remove from each diagram of this subset 
the loop propagator $i$ we obtain a set of tree diagrams.
After removing multiple copies of identical diagrams this set 
forms a Born partial amplitude which we denote by $A^{(0)}_{i}$.
After integration, the soft and collinear poles of a primitive one-loop amplitude with massless partons
are given by \cite{Giele:1992vf,Kunszt:1994mc,Catani:1997vz}
\bq
\label{ir_poles}
\lefteqn{
 S_\eps^{-1} \mu^{2\eps} A^{(1)}_{bare}
 = 
 \frac{\alpha_s}{4\pi} 
 \frac{e^{\eps \gamma_E}}{\Gamma(1-\eps)}
 \sum\limits_{i \in I_g}
} & & \\
 & &
 \left[
  \frac{2}{\eps^2}
                          \left( \frac{-2p_i\cdot p_{i+1}}{\mu^2} \right)^{-\eps}
  + \frac{2}{\eps} \left( S_i + S_{i+1} \right) 
 \right]
 A^{(0)}_i
 + {\cal O}(\eps^0).
 \nonumber 
\eq
We have multiplied the one-loop amplitude by $S_\eps^{-1} \mu^{2\eps}$, where $S_\eps = (4\pi)^\eps e^{-\eps\gamma_E}$ is the 
typical volume factor of dimensional regularisation, $\gamma_E$ is Euler's constant  and $\mu$ is the renormalisation scale.
The constants $S_i$ are given by $S_q=S_{\bar{q}}=1$ and $S_g=1/2$, the index $i$ of $S_i$ refers to the external particles.
The generalisation of eq.~(\ref{ir_poles}) to massive partons is also known.

Eq.~(\ref{ir_poles}) describes the infrared poles after integration over the loop momentum.
In this letter we give a formula which approximates the integrand of eq.~(\ref{starting_point})
in all infrared singular regions before the integration over the loop momentum.
It thus defines a counter-term for the infrared singularities which is local in the loop momentum $k$ 
and the external momenta $p_i$.

%-----------------------------------------------------------
\section{The formula}
\label{sect:formula}

In this section we present the formula for massless QCD.
The extension to massive particles is discussed at the end of this letter.
The infrared singular part can be written in unintegrated form as a soft and a collinear part:
\bq
\label{subtraction_term}
 G_{IR}^{(1)} & = & G_{soft}^{(1)} + G_{coll}^{(1)},
 \\
 G_{soft}^{(1)} & = &
 - 4 \pi \alpha_s i
 \sum\limits_{i \in I_g}
 \frac{4 p_i \cdot p_{i+1}}{k_{i-1}^2 k_i^2 k_{i+1}^2}  A^{(0)}_i,
 \nonumber  \\
 G_{coll}^{(1)} & = & 
 - 4 \pi \alpha_s i
 \sum\limits_{i\in I_g}
 (-2) 
 \left( 
 \frac{S_i g_{UV}\left(k_{i-1}^2,k_i^2,\mu_c^2\right) }{k_{i-1}^2 k_i^2}
 \right. \nonumber \\
 & & \left.
+
 \frac{S_{i+1} g_{UV}\left(k_{i}^2,k_{i+1}^2,\mu_c^2\right) }{k_{i}^2 k_{i+1}^2}
 \right)
 A^{(0)}_i.
 \nonumber
\eq
The Born partial amplitude $A^{(0)}_i$ depends on the external momenta, but not 
on the loop momentum.
The function $g_{UV}$ ensures a regular behaviour of  the collinear term in the ultraviolet region.
A possible choice is \cite{Nagy:2003qn}
\bq
 g_{UV}\left(k_{i-1}^2,k_i^2,\mu_c^2\right) & = &
 \frac{1}{2} \left( \frac{-\mu_c^2}{k_{i-1}^2-\mu_c^2} + \frac{-\mu_c^2}{k_{i}^2-\mu_c^2} \right).
\eq
$\mu_c$ is an arbitrary scale.
Integrating the soft and the collinear part we obtain
\bq
\label{integrated_ir_subtraction}
\lefteqn{
 S_\eps^{-1} \mu^{2\eps} \int \frac{d^Dk}{(2\pi)^D} G_{soft}^{(1)} = 
 \frac{\alpha_s}{4\pi} 
 \frac{e^{\eps \gamma_E}}{\Gamma(1-\eps)}
} & & \\
 & & \times
 \sum\limits_{i \in I_g}
 \frac{2}{\eps^2} 
 \left( \frac{-2p_i\cdot p_{i+1}}{\mu^2} \right)^{-\eps}
 A^{(0)}_i
 + {\cal O}(\eps),
\nonumber \\
\lefteqn{
 S_\eps^{-1} \mu^{2\eps} \int \frac{d^Dk}{(2\pi)^D} G_{coll}^{(1)} = 
 \frac{\alpha_s}{4\pi} 
 \frac{e^{\eps \gamma_E}}{\Gamma(1-\eps)}
} & & \nonumber \\
 & & \times
 \sum\limits_{i \in I_g}
 \left( S_i + S_{i+1} \right)
 \left( \frac{\mu_c^2}{\mu^2} \right)^{-\eps} \left( \frac{2}{\eps} + 2 \right)
 A^{(0)}_i
 + {\cal O}(\eps).
 \nonumber
\eq
Equation~(\ref{subtraction_term}) is the main result of this letter.
This formula approximates the integrand of a primitive one-loop QCD amplitude in all soft and collinear limits.
The approximation is given by simple scalar two- and three-point functions, multiplied by a Born
partial amplitude.
One easily observes that the integrated form in eq.~(\ref{integrated_ir_subtraction}) 
agrees in the pole terms with eq.~(\ref{ir_poles}).

%-----------------------------------------------------------
\section{Proof of the formula}
\label{sect:proof}

We first review briefly under which conditions infrared singularities occur in 
an individual Feynman diagram \cite{Kinoshita:1962ur,Nagy:2003qn,Dittmaier:2003bc}:
Soft singularities occur when a massless particle is exchanged between two on-shell particles.
With the notation as in eq.~(\ref{starting_point}) 
this corresponds to the case
\bq
 m_i=0, \;\;\; p_i^2=m_{i-1}^2, \;\;\; p_{i+1}^2=m_{i+1}^2.
\eq
In that case the propagators $(i-1)$, $i$ and $(i+1)$ are on-shell. The singularity comes from the integration region
$k \sim q_i$.
A collinear singularity occurs if a massless external on-shell particle is attached to two massless propagators.
This corresponds to
\bq
 p_i^2=0,
\;\;\;
 m_{i-1}=0,
\;\;\;
 m_{i}=0.
\eq
In that case the propagators $(i-1)$ and $i$ are on-shell. The singularity comes from the integration region
$k \sim q_i - x p_i$,
where $x$ is a real variable between 0 and 1.
In order to proof eq.~(\ref{subtraction_term})
we consider now massless QCD amplitudes.

The soft subtraction term is derived as follows:
In the case where gluon $i$ is soft, the corresponding propagator goes on-shell 
and we may replace in all Feynman diagrams which have propagator $i$ the metric tensor $g_{\mu\nu}$ of this propagator by
a polarisation sum and gauge terms:
\bq
\label{gluon_prop_replacement}
 \frac{-ig^{\mu\nu}}{k_i^2} \rightarrow 
 \frac{i}{k_i^2} \left( d^{\mu\nu}(k_i^\flat,n) 
                      - 2 \frac{k_i^{\flat \mu} n^\nu + n^\mu k_i^{\flat \nu} }{2 k_i^\flat \cdot n} \right).
\eq
Here $k_i^\flat$ denotes the on-shell limit of $k_i$ and $d^{\mu\nu}$ denotes the sum over the physical polarisations:
\bq
 d^{\mu\nu}(k,n) & = &
 \sum\limits_{\lambda} \varepsilon^\mu_{\lambda}(k,n) \varepsilon^\nu_{-\lambda}(k,n)
\nonumber \\
 & = &
 -g^{\mu\nu} + 2 \frac{k^\mu n^\nu + n^\mu k^\nu}{2 k \cdot n}.
\eq
$n^\mu$ is a light-like reference vector.
We note that self-energy diagrams are not singular in the soft limit, therefore adding them to the loop diagrams
will not change the soft limit.
With the inclusion of the self-energy diagrams and a corresponding replacement as in eq.~(\ref{gluon_prop_replacement})
the contribution from the polarisation sum in eq.~(\ref{gluon_prop_replacement})
makes up a tree-level partial amplitude, 
where two gluons with momenta $k_i^\flat$ and $-k_i^\flat$ have been inserted between the
external legs $i$ and $i+1$.
In the soft limit this tree-level partial amplitude is given by two eikonal factors times the tree-level
partial amplitude without these two additional gluons:
\bq
\label{two_eikonal}
 \left( i g \frac{p_i^\mu}{p_i \cdot k_i^\flat} \right) 
 g_{\mu\nu} 
 \left( i g \frac{p_{i+1}^\nu}{p_{i+1}\cdot(-k_i^\flat)} \right) A^{(0)}_i.
\eq
In the soft limit we may replace $2 p_i \cdot k_i^\flat$ by $k_{i-1}^2$ and
$2 p_{i+1}\cdot(-k_i^\flat)$ by $k_{i+1}^2$.
Eq.~(\ref{two_eikonal}) then leads to the soft part of eq.~(\ref{subtraction_term}).
The terms with $k_i^{\flat \mu} n^\nu$ and $n^\mu k_i^{\flat \nu}$ in eq.~(\ref{gluon_prop_replacement})
vanish for the sum of all diagrams due to gauge invariance.

The collinear subtraction term is derived as follows:
We have to consider diagrams, where two adjacent propagators in the loop go on-shell with a massless external leg
in between.
The cases where an external gluons splits into a 
ghost-antighost-pair or into a quark-antiquark-pair are in the collinear limit 
not singular enough to yield a divergence after integration.
Therefore we are left with the case where an external quark splits into a quark-gluon pair
and the case where an external gluon splits into two gluons.
Let us first consider the $q \rightarrow q g$ splitting.
In Feynman gauge one can show that only the longitudinal polarisation of the gluon
contributes to the collinear limit.
The same holds true for the $g \rightarrow gg$ splitting.
In this case the collinear limit receives contributions when one of the two gluons in the loop
carries a longitudinal polarisation (but not both).
The external gluon has of course physical transverse polarisation.
We can now use gauge-invariance to turn the sum of all diagrams with a collinear singularity in the propagators
$(i-1)$ and $i$ into the simple form of eq.~(\ref{subtraction_term}).
The contraction of a longitudinal polarisation into a gauge-invariant set of diagrams 
yields zero.
The set of collinear divergent diagrams forms an almost gauge-invariant set of
diagrams. There is only one diagram missing, where the longitudinal polarised gluon couples directly
to the other parton.
This is a self-energy insertion on an external line, which by definition is absent from
the amputated one-loop amplitude.
We can now turn the argument around and replace the sum of collinear singular diagrams by the negative
of the self-energy insertion on the external line.
The self-energy insertions on the external lines introduce a spurious $1/p_i^2$-singularity.
In order to calculate the singular part of the self-energies we regulate this spurious singularity
by allowing $p_i^2$ slightly off-shell, but keeping $k_{i-1}$ and $k_i$ on-shell and imposing momentum conservation.
We can use the same parametrisation as in the real emission case:
\bq
 k_{i-1} & = & x p + k_\perp - \frac{k_\perp^2}{x} \frac{n}{(2p\cdot n)},
 \nonumber \\
 -k_{i} & = & (1-x) p - k_\perp - \frac{k_\perp^2}{(1-x)} \frac{n}{(2p\cdot n)},
\eq
with $p^2=n^2=0$ and $2 p\cdot k_\perp = 2 n \cdot k_\perp = 0$.
The singular part of the self-energies with one longitudinal polarised gluon is proportional to 
\bq
 P_{q\rightarrow q g}^{long} & = & 
 - \frac{2}{2k_{i-1}\cdot k_i} \left( - \frac{2}{1-x} + 2 \right) p\!\!\!/,
 \\
 P_{g\rightarrow g g}^{long} & = & 
 - \frac{2}{2k_{i-1}\cdot k_i} 
 \left( - \frac{2}{x} - \frac{2}{1-x} + 2 \right) d^{\mu\nu}(p,n).
 \nonumber
\eq
The terms with $2/x$ and $2/(1-x)$ correspond to soft singularities and have already been subtracted out 
with the soft subtraction term $G_{soft}^{(1)}$. 
In the collinear limit we therefore just have to subtract out the terms, which are non-singular in the
soft limit. These terms are independent of $x$ and lead to the collinear part of eq.~(\ref{subtraction_term}).

A few remarks are in order: Eq.~(\ref{ir_poles}) is usually stated for the renormalised one-loop amplitude.
\bq
 {\cal A}^{(1)}_{ren}
 & = &
 \frac{\alpha_s}{4\pi} 
 \frac{e^{\eps \gamma_E}}{\Gamma(1-\eps)}
 \left[
 \sum\limits_{\mbox{\scriptsize pairs }(i,j)} \frac{2 {\bf T}_i {\bf T}_j}{\eps^2}  
                          \left( \frac{-2p_i \cdot p_{i+1}}{\mu^2} \right)^{-\eps}
 \right. \nonumber \\
 & & \left.
 - \sum\limits_i \frac{\gamma_i}{\eps} 
\right]
 {\cal A}^{(0)}
 + {\cal O}(\eps^0).
\eq
The colour charge operators are given by ${\bf T}_q=T^a_{ij}$,
${\bf T}_{\bar{q}}=-T^a_{ji}$ and ${\bf T}_g=if^{bac}$ for final state particles.
The constants $\gamma_i$ are given by $\gamma_q=\gamma_{\bar{q}}=3 C_F/2$ and $\gamma_g=\beta_0/2$ with 
$\beta_0 = 11 C_A/3 - 4 T_R N_f/3$.
The renormalised one-loop amplitude with $n_g$ gluons, $n_q$ quarks and $n_{\bar{q}}$ antiquarks
is related to the bare amplitude by
\bq
\label{LSZ}
 {\cal A}_{ren}(p_1,...,p_n,\alpha_s)
 & = &
 \left(Z_2^{1/2} \right)^{n_q+n_{\bar q}} \left( Z_3^{1/2} \right)^{n_g}
 \\
& &  
 \times {\cal A}_{bare}\left(p_1,...,p_n,Z_g^2 S_\eps^{-1}\mu^{2\eps} \alpha_s\right).
 \nonumber
\eq
$Z_g$ is coupling renormalisation constant 
\bq
 Z_g & = & 1 + \frac{\alpha_s}{4\pi} \left( - \frac{\beta_0}{2} \right)  \frac{1}{\eps} + {\cal O}(\alpha_s^2).
\eq
$Z_2$ is the quark field renormalisation constant and $Z_3$ is the gluon field renormalisation constant.
The LSZ reduction formula instructs us to take for the field renormalisation constants the residue
of the propagators at the pole.
In dimensional regularisation this residue is $1$ for massless particles and therefore the field renormalisation
constants are often omitted from eq.~(\ref{LSZ}).
However $Z_2=Z_3=1$ is due to a cancellation between ultraviolet and infrared divergences.
In Feynman gauge we have
\bq
 Z_{2} & = & 1 + \frac{\alpha_s}{4\pi} C_F \left( \frac{1}{\eps_{IR}} - \frac{1}{\eps_{UV}} \right) + {\cal O}(\alpha_s^2),
 \\
 Z_{3} & = & 1 + \frac{\alpha_s}{4\pi} \left( 2 C_A -\beta_0 \right) \left( \frac{1}{\eps_{IR}} - \frac{1}{\eps_{UV}} \right) + {\cal O}(\alpha_s^2).
 \nonumber 
\eq
Here we indicated explicitly the origin of the $1/\eps$-poles.
Using $Z_2=Z_3=1$ mixes therefore poles of ultraviolet and infrared origin.
There are different variants of dimensional regularisation: conventional dimensional regularisation,
the 't Hooft-Veltman scheme and four-dimensional schemes.
The soft and collinear subtraction terms in eq.~(\ref{subtraction_term})
are independent of the variant of dimensional regularisation.
The infrared poles obtained from the phase space integration of the real emission amplitude depend on the other hand
on the scheme of dimensional regularisation.
Again the solution is due to the field renormalisation constants which mix scheme-dependent terms of ultraviolet and
infrared origin \cite{Harris:2002md}.
The scheme-dependence of the bare one-loop amplitude is entirely of ultraviolet origin.

%-----------------------------------------------------------
\section{Generalisation to massive partons}
\label{sect:massive}

In this section we present the generalisation to massive QCD.
This is in particular relevant to top quark physics. There are only a few modifications
necessary with respect to the massless case.
The modification for the unintegrated soft subtraction term is straightforward:
\bq
\lefteqn{
 G_{soft}^{(1)} = 
} & & \\
 & & 
 - 4 \pi \alpha_s i
 \sum\limits_{i \in I_g}
 \frac{4 p_i \cdot p_{i+1}}{\left( k_{i-1}^2 - m_{i-1}^2\right) k_i^2 \left( k_{i+1}^2 - m_{i+1}^2\right)}
 A^{(0)}_i.
 \nonumber 
\eq
As before, the sum runs over all particles in the loop which are gluons.
Integrating the soft subtraction term we have to distinguish whether the masses $m_{i-1}$ and $m_{i+1}$ 
are zero or not.
The result can be written as
\bq
\lefteqn{
 S_\eps^{-1} \mu^{2\eps}  \int \frac{d^Dk}{(2\pi)^D}
 G_{soft}^{(1)} = 
 \frac{\alpha_s}{4\pi} 
 \frac{e^{\eps \gamma_E}}{\Gamma(1-\eps)}
 \sum\limits_{i\in I_g}
} & & \\
 & & \times
 C\left((p_i+p_{i+1})^2,m_{i-1}^2,m_{i+1}^2,\mu^2\right)
 A^{(0)}_i
 + {\cal O}(\eps),
 \nonumber
\eq
where the function $C(s,m_1^2,m_2^2,\mu^2)$ is given for the four different cases by \cite{Dittmaier:2003bc,Ellis:2007qk}
\bq
%\lefteqn{
C\left(s,0,0,\mu^2\right) 
 & = & 
 \frac{2}{\eps^2} 
 \left( \frac{-s}{\mu^2} \right)^{-\eps},
\\
C\left(s,0,m^2,\mu^2\right)
  & = & 
C\left(s,m^2,0,\mu^2\right),
 \nonumber 
% } & &
% \\
\eq
\bq
\lefteqn{
C\left(s,m^2,0,\mu^2\right) 
 =
 \left( \frac{m^2}{\mu^2} \right)^{-\eps}
 \left[ \frac{1}{\eps^2} 
 + \frac{2}{\eps} \ln\left(\frac{m^2}{m^2-s}\right) 
 \right.
} & &
 \nonumber \\
 & & 
 \left.
 + \frac{\pi^2}{6}
 + \ln^2\left(\frac{m^2}{m^2-s}\right) - 2 \; \mbox{Li}_2\left(\frac{-s}{m^2-s}\right) \right],
 \nonumber \\
%\lefteqn{
%C\left(s,0,m^2,\mu^2\right)
% = 
%C\left(s,m^2,0,\mu^2\right),
% } & & \nonumber \\
\lefteqn{
C\left(s,m_1^2,m_2^2,\mu^2\right) 
 =  
 \frac{2x(s-m_1^2-m_2^2)}{m_1m_2(1-x^2)} 
 \left\{ 
                \left[ -\frac{1}{\eps} 
                       - \frac{1}{2} \ln(x) 
\right. \right.
} & & \nonumber \\
 & &
 \left. \left. 
                       + 2 \ln(1-x^2) + \ln\left(\frac{m_1m_2}{\mu^2}\right) \right]
         \ln(x) 
         - \frac{\pi^2}{6} + \mbox{Li}_2(x^2) 
 \right. \nonumber \\
 & & \left.
         + \frac{1}{2} \ln^2\left(\frac{m_1}{m_2}\right)
         + \mbox{Li}_2\left(1-x\frac{m_1}{m_2}\right) + \mbox{Li}_2\left(1-x\frac{m_2}{m_1}\right)
 \right\},
 \nonumber 
\eq
with
\bq
 x & = & - \frac{1-\chi}{1+\chi},
 \;\;\;\;\;
 \chi = \sqrt{1-\frac{4m_1m_2}{s-\left(m_1-m_2\right)^2} }.
\eq
The modification for the collinear subtraction term is even simpler:
There is no collinear singularity if an external quark or antiquark is massive.
It suffices therefore to define $S_Q=S_{\bar{Q}}=0$ for a massive quark or antiquark.

%-----------------------------------------------------------
\section{Conclusions}
\label{sect:conclusions}

In this letter we have shown that the infrared singular part of the integrand of a primitive one-loop
QCD amplitude is given by simple scalar two- and three-point functions, multiplied by a Born partial amplitude.
The immediate application is to use this form as a subtraction term for the numerical integration of a one-loop
amplitude.
In view of refs.~\cite{Becher:2009cu,Gardi:2009qi,Ferroglia:2009ep} we do not exclude the possibility that this method generalises to higher loops.

%-----------------------------------------------------------

\begin{appendix}

\section{Primitive amplitudes}

In this appendix we include a brief summary on primitive one-loop amplitudes.
In order to construct a primitive one-loop amplitude one starts 
to draw all possible planar one-loop diagrams with
a fixed cyclic order of the external legs, subject to the constraint that each fermion line is either
only left-moving or only right-moving. We call a fermion line ``left-moving'', if following the
arrow of the fermion line the loop is to the right.
In an amplitude with $r$ external quark-antiquark pairs $p$ of these pairs can be left-moving
The remaining $(r-p)$ pairs are then right-moving.
\begin{figure}
\begin{center}
\includegraphics[bb= 165 510 445 713, width=0.4\textwidth]{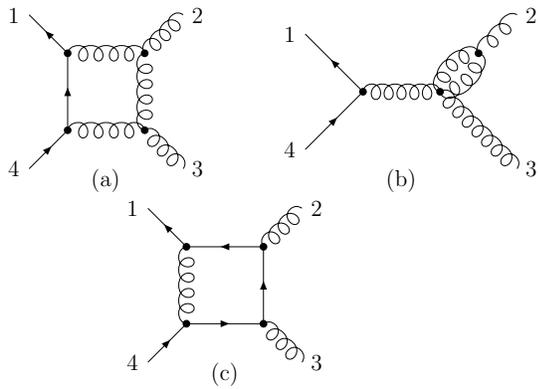}
\caption{\label{figure_example_qqgg}
Examples of diagrams: Diagrams (a) and (b) contribute to the left-moving primitive amplitude,
while diagram (c) contributes to the right-moving amplitude.
}
\end{center}
\end{figure}
\begin{figure}
\begin{center}
\includegraphics[bb= 125 423 492 695, width=0.4\textwidth]{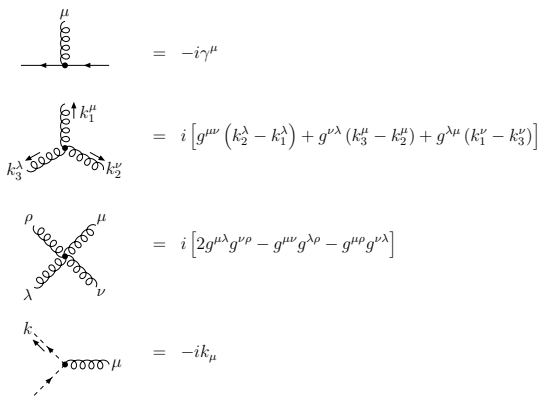}
\caption{\label{figure_vertices}
Colour-ordered Feynman rules.
}
\end{center}
\end{figure}
As an example we consider primitive $\bar{q}ggq$-amplitudes with the cyclic order $(1_{\bar{q}},2{g},3_{g},4_q)$.
Fig.~\ref{figure_example_qqgg} shows some diagrams as examples.
Diagrams (a) and (b) contribute to the primitive amplitude, where the quark line is left-moving,
diagram (c) however contributed to the primitive amplitude, where the quark line is right-moving.
We would like to point out that diagram (b) has to be included. In an analytic calculation this diagram
is often discarded, because it yields zero within dimensional regularisation.
However this zero is obtained from a cancellation between an ultraviolet divergence and a collinear divergence.
In a numerical calculation we have to keep this diagram in order not to spoil the local structure.

A diagram is translated to a formula with the help of the following Feynman rules:
The propagators for a quark, gluon and ghost particle are
\bq
 i\frac{k\!\!\!/+m}{k^2-m^2},
\;\;\;\;\;\;
 \frac{-ig^{\mu\nu}}{k^2},
\;\;\;\;\;\;
 \frac{i}{k^2},
\eq
respectively.
The Feynman rules for the vertices are listed in fig.~\ref{figure_vertices}.

In the main part of the paper we made use of the Ward identity for cyclic ordered amplitudes.
The Ward identity states if we replace in the amplitude
the polarisation vector of one external gluon by its momentum then
we obtain zero. A proof can be found in the textbook by Peskin and Schroeder \cite{Peskin}.

\end{appendix}

%-----------------------------------------------------------

\bibliography{/home/stefanw/notes/biblio}

\end{document}